\documentclass[apj]{emulateapj}
\usepackage{amsmath}

\newcommand{\etal}{{et al.~}}

\newcommand{\kpch}{\>h^{-1}{\rm {kpc}}}

\newcommand{\kmsmpc}{\>{\rm km}\,{\rm s}^{-1}\,{\rm Mpc}^{-1}}

\newcommand{\rmag}{\>^{0.1}{\rm M}_r-5\log h}

\usepackage{color}

\shorttitle{Connections between galaxy mergers and starburst}
\shortauthors{Luo et al.}

\begin{document}


\title{Connections between galaxy mergers and Starburst: evidence from local
  Universe}

\author{Wentao Luo\altaffilmark{1,2}, Xiaohu Yang\altaffilmark{1,3},
  Youcai Zhang\altaffilmark{1,2}}

\altaffiltext{1} {Key Lab for Research in Galaxies and Cosmology, Shanghai
  Astronomical Observatory, Nandan Road 80, Shanghai 200030, China E-mail:
  walt@shao.ac.cn}

\altaffiltext{2} {University of the Chinese Academy of Sciences, 19A
Yuquan
  Road, Beijing 100049, China}

\altaffiltext{3} {Center for Astronomy and Astrophysics, Shanghai
Jiao Tong
  University, 800 Dongchuan Road, Shanghai 200240, China E-mail:
  xyang@sjtu.edu.cn }

\begin{abstract}
  Major mergers and interactions between gas-rich galaxies with
  comparable masses are thought to be the main triggers of starburst.
  In this work, we study, for a large stellar mass range, the
  interaction rate of the starburst galaxies in the local universe. We
  focus independently on central and satellite star forming galaxies
  extracted from the Sloan Digital Sky Survey. Here the
  starburst galaxies are selected in the star formation rate (SFR)
  stellar mass plane with SFR five times larger than the median value
  found for ``star forming'' galaxies of the same stellar mass.
  Through visual inspection of their images together with close
  companions determined using spectroscopic redshifts, we find that
  $\sim$50\% of the ``starburst'' populations show evident merger
  features, i.e., tidal tails, bridges between galaxies, double cores
  and close companions. In contrast, in the control sample we
  selected from the normal star forming galaxies, only
  $\sim$19\% of galaxies are associated with evident mergers.
  The interaction rates may increase by $\sim$5\% for the
  starburst sample and 2\% for the control sample if close companions
  determined using photometric redshifts are considered.  The contrast
  of the merger rate between the two samples strengthens the hypothesis that
  mergers and interactions are indeed the main causes of starburst.
\end{abstract}



\section{INTRODUCTION}

Amongst various mechanisms that trigger starburst in galaxies,
interactions and merging processes between gas-rich galaxies are
regarded as the most efficient ones (Larson \& Tinsley 1978;
Kennicutt 1998; Elmegreen 2011).  Starbursts are typically confined
within the galaxy, and can only be triggered by the accumulation,
over a short time period, of a large amount of cold gas in a small
region. The most efficient way of achieving this is through the
merging of gas-rich galaxies, where the gas can be strongly
compressed and concentrated by tidal interactions (e.g. Mo et al.
2010).  Star formation enhancement has been measured for merging
systems over a large sample of galaxies. A factor of two to three
enhancement is found on average (e.g., Kennicutt 1998), with a large
scatter ranging from zero for dry mergers between gas-poor,
bulge-dominated systems (e.g., van Dokkum 2005) to $\sim 10-100$ for
extreme cases (e.g. Bushouse et al 1987; Telesco et al 1988).

Moreover, observational studies have also demonstrated that many
starburst galaxies may have experienced gaseous merging and have
"peculiar" morphology (e.g., Larson \& Tinsley 1978; Arp 1995).  By
performing photometric measurements on a \emph{Hubble Space
Telescope} snapshot imaging survey sample of 97 ultra-luminous
infraRed galaxies (ULIRGs), e.g., with $L_{\rm
IR}>10^{12}L_{\odot}$, at redshift $\sim 0.1$, Cui et al. (2001)
found that $\sim$18\% of ULIRGs have multiple nuclei and 39\% have
double nuclei.  Chen et al. (2010) selected 54 ULIRGs from the Sloan
Digital Sky Survey(SDSS) and found that $\sim 42\%$ of them are
merging systems. Apart from the ULIRGs at low redshift (e.g.,
Sanders et al. 1988; Melnick \& Mirabel 1990; Murphy et al. 1996),
the majority of ULIRGs at high redshift also exhibit merger features
(e.g., Kartaltepe et al. 2012).

Numerical studies of merging galaxies (e.g., Mihos \& Hernquist
1994; Springel 2000; Tissera et al. 2002; Meza et al. 2003; Kapferer
et al. 2005; Cox et al. 2006; Di Matteo et al. 2008) showed that
merging processes do trigger star formation. However the intensity
may depend on the initial disk stability, the availability of gas in
the interacting galaxies, etc.  Di Matteo et al. (2007) analyzed
several hundred simulations of galaxy collisions and compared their
star formation rates (SFR) to isolated cases.  In their simulations
only $\sim$17\% of the mergers trigger SFR 10 times larger than for
isolated galaxies, and half of the collisions have less than a
factor of four enhancement in SFR. In addition, Cox et al. (2008)
pointed out that only major mergers (galaxy pairs with comparable
stellar mass) can significantly enhance the SFRs and trigger
starbursts.

In this work, we expand those previous observational studies of
ULIRGs to less luminous starburst galaxies with $r$-band luminosity
range $10^{7}- 10^{12}L_{\odot}$, using SDSS data (York et al 2000).
We study the fractions of starburst galaxies with merger features
for central and satellite galaxies respectively. The Letter is
organized as follows: Section 2 describes the data we used, the main
results are presented in Section 3, and the summary is given in
Section 4. Throughout the Letter we apply a cosmology with
$\Omega_m=0.27, \Omega_{\Lambda}=0.73$ and $h=H_0/(100
\kmsmpc)=0.73$.

\section{Data Selection}

In this section, we describe the data selection, the definition of
starburst galaxies and the morphological classification.

\begin{figure*}
\plotone{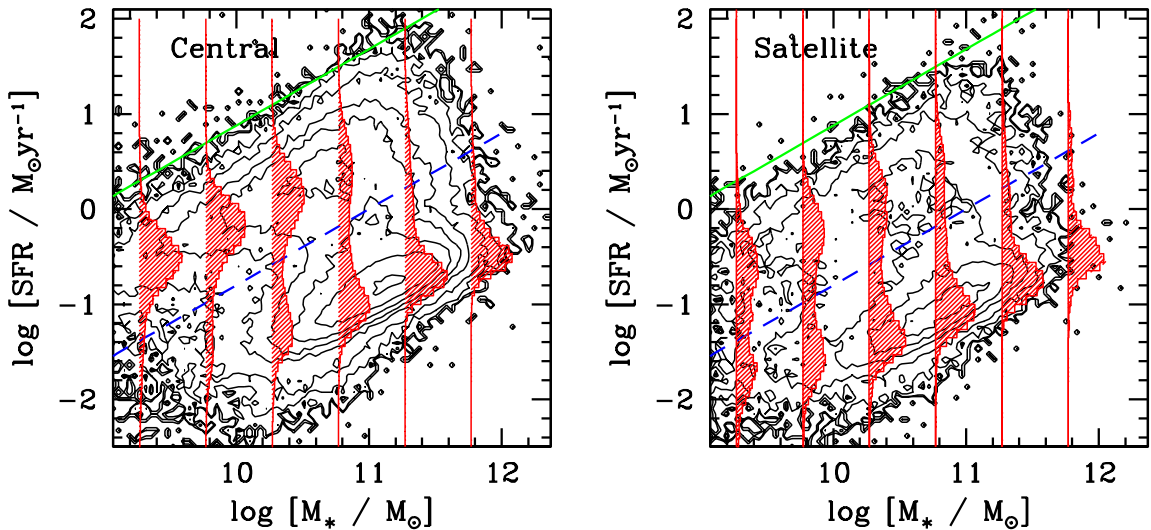} \caption{SFR vs. stellar mass plot for central
(left panel) and
  satellite (right panel) galaxies. The solid and dashed lines in each
  panel denote the selection criteria of our "starburst" galaxies:
  $\log {\rm SFR}= 0.8\times (\log M_* - 8.9)$ and star forming
  galaxies: $\log {\rm SFR}= 0.8\times (\log M_* - 11.0)$,
  respectively.}\label{fig:SFR}
\end{figure*}

\subsection{SDSS Galaxy and Group Catalogs}

To probe the interaction rates of `starburst' galaxies in the local
universe, we need to have a fair and complete sample of galaxies
with various information, e.g., images, redshifts, SFRs, etc.  For
this purpose, a galaxy sample is selected from the New York
University Value-Added Galaxy catalog (NYU-VAGC; Blanton \etal 2005)
constructed from the SDSS Data Release 7 (DR7; Abazajian \etal
2009). We have selected in the NYU-VAGC Main Galaxy Sample all
galaxies with an extinction-corrected apparent magnitude brighter
than $r=17.72$, with redshifts in the range $0.01 \leq z \leq 0.20$,
and with a redshift completeness ${\cal
  C}_z > 0.7$.  Here the redshift completeness is defined as the
average percentage of galaxies that have spectroscopic redshifts in
their local sky coverage. The resulting galaxy sample contains a
total of $599,301 $ galaxies with SDSS spectroscopic redshifts in a
sky coverage of 7748 $deg^2$. The SFRs and the stellar masses of
individual galaxies are obtained from the MPA-JHU DR7 spectrum
measurements\footnote{See
    http://www.mpa-garching.mpg.de/SDSS/DR7/}.  In this data release,
stellar masses are estimated based on their $z$ band absolute
magnitudes and a model mass-to-light ratio (Kauffmann et al. 2003;
Salim et al.  2007).  The SFRs are computed using the photometry and
emission lines as in Brinchmann et al. (2004; hereafter B04)
and Salim et al. (2007).  For star forming galaxies with strong
emission lines, the SFRs are estimated by fitting different emission
lines in the galaxy spectra. For active galactic nuclei and galaxies
with weak emission lines, SFRs are estimated from the photometries.

In order to check whether or not the interaction rates of starburst
galaxies depend on their larger scale environments, e.g., host
halos, we divide our sample into central and satellite galaxies.  We
make this distinction by using the related SDSS group catalog
(specified as modelA)\footnote{ See
http://gax.shao.ac.cn/data/Group.html} constructed by Yang \etal
(2007, hereafter Y07).  The group finder used in Y07 starts with an
assumed mass-to-light ratio to assign a tentative mass to each
potential group.  This mass is used to estimate the size and
velocity dispersion of the underlying halo that hosts the group,
which in turn is used to determine group membership (in redshift
space). The most massive group member is identified as the central
galaxy (472,504 in total), while all the other group members are
given the status of satellite galaxies (126,797 in total).

\begin{figure*}
\centering
\plotone{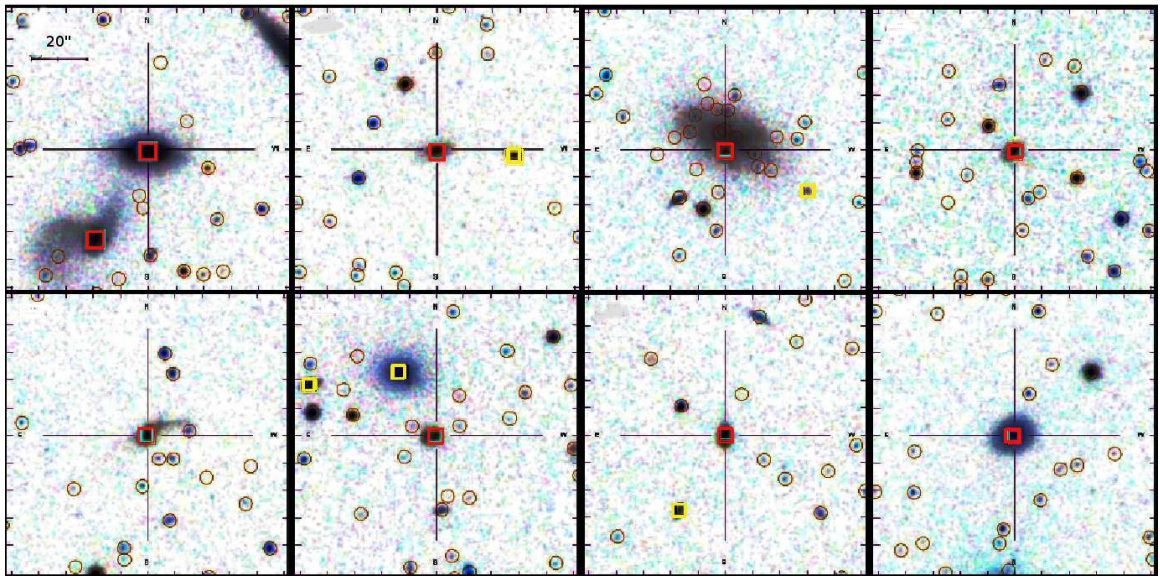}
\caption{Typical images of central galaxies (upper panels) and
  satellite galaxies (lower panels).  These galaxies are classified
  into different categories, as displayed from left to right: Type Ia,
  Type Ib, Type Ic, and Type II. The red squares denote galaxies with
  spectroscopic redshifts. Close companions of Type Ib and Type Ic
  galaxies, with photometric redshifts only, are marked by the yellow
  squares. The circles are the photometric objects detected by SDSS
  photo pipeline.
   }\label{fig:image}
\end{figure*}

Next comes the definition of our "starburst" galaxies.  As modeled
in Yang et al. (2013), the distribution of SFRs for central galaxies
shows a bimodal feature.  For a given stellar mass $M_\ast$ (or halo
mass $M_h$), the central galaxies appear to be composed of two
distinctive populations, one with high SFRs (the ``star forming''
population) and the other with SFRs that are more than 10 times
smaller (the ``quenched'' population).  In the left panel of Figure
\ref{fig:SFR}, we reproduce the figure showing the number density
distributions of `volume-limited' samples of central galaxies in the
$\log {\rm SFR}$ versus  $\log M_{\ast}$ plane using contours of
different levels (see Yang et al.  2013 for details).  The vertical
shaded histograms show the distributions of galaxies within given
logarithmic stellar mass bins of $\pm 0.2$ dex width.  According to
their bimodal feature, we can roughly separate galaxies into star
forming and quenched branches by $\log {\rm SFR} = 0.8\times(\log
M_{\ast} - 11.0)$, which is shown in the plot as the dashed line.
Similar distributions are obtained for satellite galaxies which are
shown in the right panel of Figure \ref{fig:SFR}. Compared to the
central galaxies, the satellite galaxies show a similar bimodal
feature but with different ratios as a function of stellar mass:
there are many more quenched low-mass satellite galaxies than
central galaxies (e.g., a few percent of quenched central galaxies
versus $\sim 50\%$ quenched satellite galaxies with stellar mass
$\sim 10^{9} M_{\odot}$).

The definition of our "starburst" galaxies is set as follows,
\begin{equation}
  \log {\rm SFR}\ge 0.8\times(\log M_{\ast} - 8.9)\,.
\label{eq:starburst}
\end{equation}
Thus defined, our "starburst" galaxies have SFRs that are $\sim$5
times larger than the median SFRs of the ``star forming'' galaxies
as a function of stellar mass. Such "starburst" definition is
similar to that introduced by Rodighiero et al. (2011).  The above
selection criteria is illustrated in Figure\ref{fig:SFR} using solid
lines. Note here that the factor five is arbitrary. However, setting
this factor to 10 or 2 would result in too few or too many galaxies
for our visual inspection. According to our definition, a total of
1455 centrals and 404 satellites fall into our "starburst" galaxy
categories. Once compared to the total number of central (472,504)
and satellite (126,797) galaxies, the fractions of starburst
satellites we selected are $\sim$0.3\% in both cases.

Apart from our definition of the "starburst" galaxies which are
 based on SFR and $M_{\ast}$, there are other definitions
of `starburst' galaxies according to line features (e.g., Kewley et
al.  2001; Kauffmann et al. 2003; B04; Lee et al. 2007; 2009; Bolton
et al. 2012).  According to the definition of Bolton et al.  (2012),
star forming galaxies with ${\rm H}_{\alpha}$ equivalent width
larger than $50\AA$ are defined as "starburst" galaxies. Using this
definition we can obtain 13460 central and 1649 satellite
"starburst" galaxies, $\sim$10 times
 more than ours. The distribution of their
"starburst" galaxies in the $\log {\rm SFR}$ versus  $\log M_{\ast}$
plane spread all over the "star forming" branch.  Still they tend to
have somewhat enhanced SFRs with respect to the median SFR as a
function of stellar mass. About $81\%$ of our "starburst" galaxies
fall into the categories of Bolton et al. (2012).  Thus our
starburst galaxies are a small selection among theirs with very high
SFRs.

In order to assess a causal relationship between interactions and
starbursts, we constructed a control sample of "normal" star forming
galaxies. This control sample contains the same number of randomly
selected star forming central and satellite galaxies with respective
redshift and stellar mass distributions similar to those of the
starburst galaxy sample.

\begin{table*}[]
\begin{center}
  \caption{Fractions of galaxies being classified in different
    merger or interaction types.}
\begin{tabular}{lccccc}
\hline
Type~~ & \multicolumn{2}{c}{Centrals (1455)} & &
\multicolumn{2}{c}{Satellites (404)} \\
\cline{2-3}  \cline{5-6}
  &  starburst (major/wet) & control  (major/wet)  & &starburst
(major/wet)  & control  (major/wet) \\
\hline
Ia   & 50.6$\pm$0.6\% (--/--) & 19.1$\pm$0.7\% (--/--)
&& 48.8$\pm$2.5\% (--/--) & 17.1$\pm$1.8\% (--/--)  \\
(Ia pairs)  & 21.1$\pm$0.4\% (63.2\%/94.8\%) & 11.3$\pm$0.6\%
(91.1\%/84.4\%)
&& 12.9$\pm$0.8\% (65.5\%/94.6\%) & 5.9$\pm$0.5\% (57.9\%/89.5\%)  \\
Ib   & 2.9$\pm$0.4\% (47.6\%/52.4\%) & 0.9$\pm$0.1\% (35.7\%/35.7\%)
&& 2.5$\pm$0.8\% (60.0\%/50.0\%) & 0.7$\pm$0.4\% (0.0\%/66.7\%)\\
Ic   & 1.8$\pm$0.3\% (48.1\%/25.9\%)  & 0.8$\pm$0.2\%
(45.5\%/36.3\%)
&& 2.5$\pm$0.8\% (40.0\%/40.0\%) & 0.0$\pm$0.0\% (--/--)\\
II   & 44.7$\pm$0.5\% (--/--) & 79.1$\pm$0.8\% (--/--) &&
46.0$\pm$2.4\% (--/--) & 82.2$\pm$2.1\% (--/--)\\
\hline

\end{tabular}\label{tab:frac}
\end{center}
\end{table*}

\subsection{Morphological Classification}

In order to proceed to the eyeball checking of both our starburst
and normal star forming galaxy samples, we have downloaded images
synthesized from g, r, i band (Lupton et al 2004) of $1455\times 2$
central and $404\times 2$ satellite galaxies from the SDSS website.
We have classified these galaxies into different categories
following the criteria used by Bridge et al. (2007) for images in
the Spitzer First Look Survey. Images showing tidal features, e.g.
tidal tails, double cores and bridges between galaxies etc.  (see
the left panel images of Fig. \ref{fig:image}) are defined as Type
Ia. In addition, galaxies with `dynamical' close companions are
classified as Type Ia as well. Here dynamical close companions are
defined as galaxies having velocity differences less than 500${\rm
km/s}$ and projected distance smaller than $20\kpch$ (e.g. Patton et
al. 2000). These Type Ias can be treated as merger associated
galaxies.  As the SDSS spectroscopic observation suffer from the
fiber collision effect, no galaxy pair can be observed with
separation smaller than 55" in one observation plate, except in the
overlapped plates (Mesa et al 2013). Thus, for completeness,
photometric redshifts are used as well. Galaxies without close
spectroscopic companions but with close photometric companions are
classified as Type Ib (Ic).  Photometric close companions are here
defined as follows. (1) They have consistent photometric redshifts
as the reference galaxy (i.e., within 1-$\sigma$ photometric
redshift error of the companion galaxy, on average at about 0.023).
(2) Their projected distances are smaller than $20\kpch$. (3) their
$r$ band apparent magnitudes are brighter than 20.0 so that very
minor merger cases (with stellar mass ratio $\la 1/10$) are
excluded.  We make use of two sets of photometric redshifts provided
by the SDSS webserver. The first set is based on the Random Forests
(RF) method proposed by Carliles et al. (2010), where the
photometric redshifts are determined using a set of optimal decision
trees built on subsets of spectroscopic samples. The other set is
generated based on a hierarchical indexing structures method
(KD-trees; see Csabai et al. 2003 and references therein). In Type
Ib, the photometric redshifts of the companions measured in both RF
and KD-tree methods are consistent with the spectroscopic redshift
of the reference galaxy.  Type Ic has at least one close companion
but with only either RF or KD-tree redshift consistent with the
reference galaxy.  Thus defined, the Type Ib+Ic can be treated as
the possible merger associated galaxies, as the photometric
redshifts are not very reliable observables for this purpose. The
remaining galaxies are all considered to be Type II meaning that
neither mergers nor close companions except those with stellar mass
ratio $\la 1/10$ are found. Thus, we can take the fraction of Type
Ia as the lower limit and fraction of Type Ia+Ib+Ic as the upper
limit of the merger rate of our galaxy samples.  As an illustration,
we show in the upper and lower panels of Figure \ref{fig:image} the
selected sample images of central and satellite galaxies.

\begin{figure*}
\centering
\includegraphics[width=8cm,height=15cm]{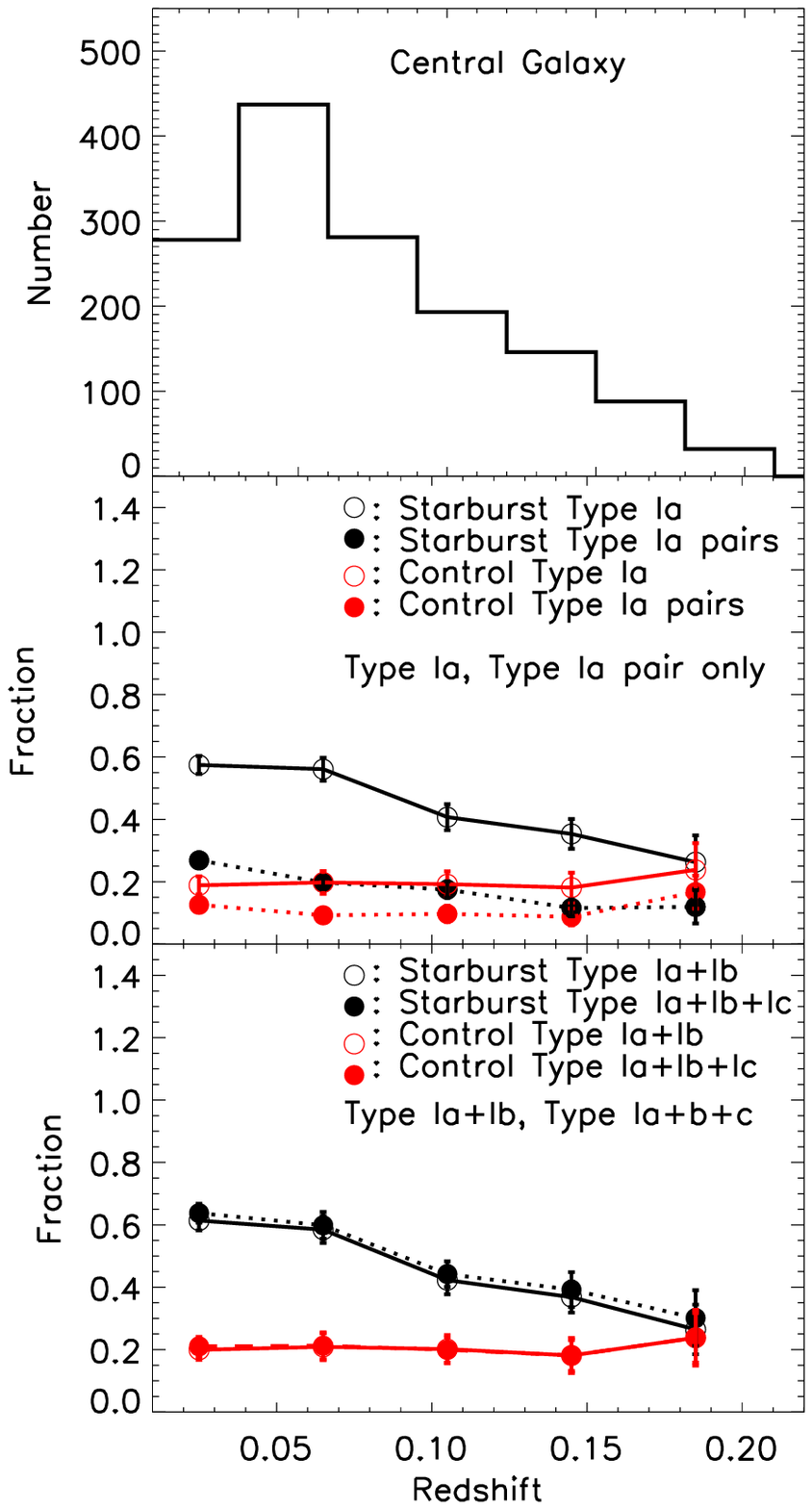}
\includegraphics[width=8cm,height=15cm]{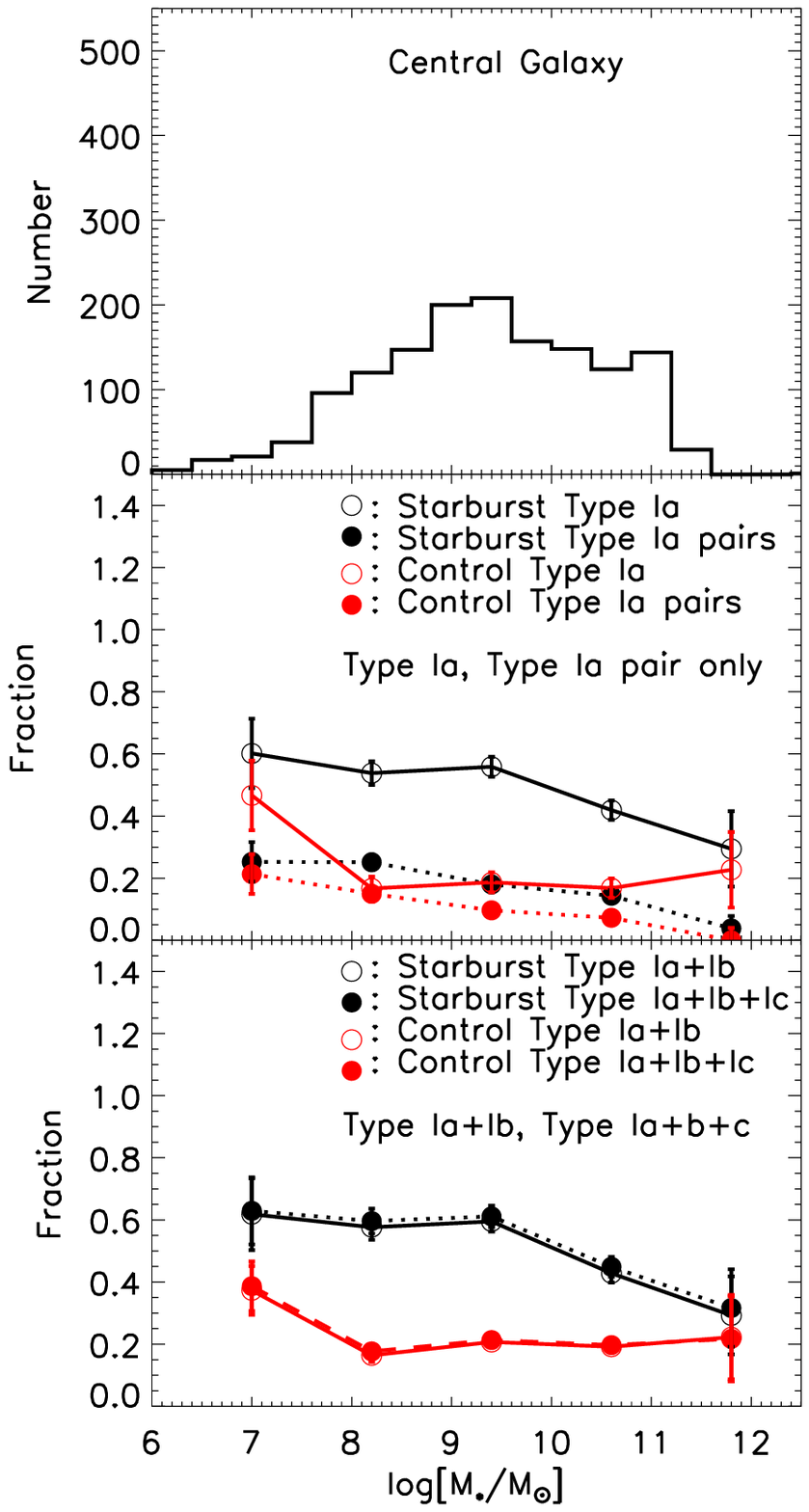}
\caption{Shown in the upper left and right panels are the redshift
  and stellar mass distributions of our starburst central galaxies.
  The middle left and right panels are the fractions of different
  morphology types (Type Ia, Type Ia with spectroscopic close
  companions) as a function of redshift and stellar mass.  The lower
  panels are similar to the middle panels for Type Ia+Ib and Type
  Ia+Ib+Ic. }
  \label{fig:cent}
\end{figure*}

\begin{figure*}
\centering
\includegraphics[width=8cm,height=15cm]{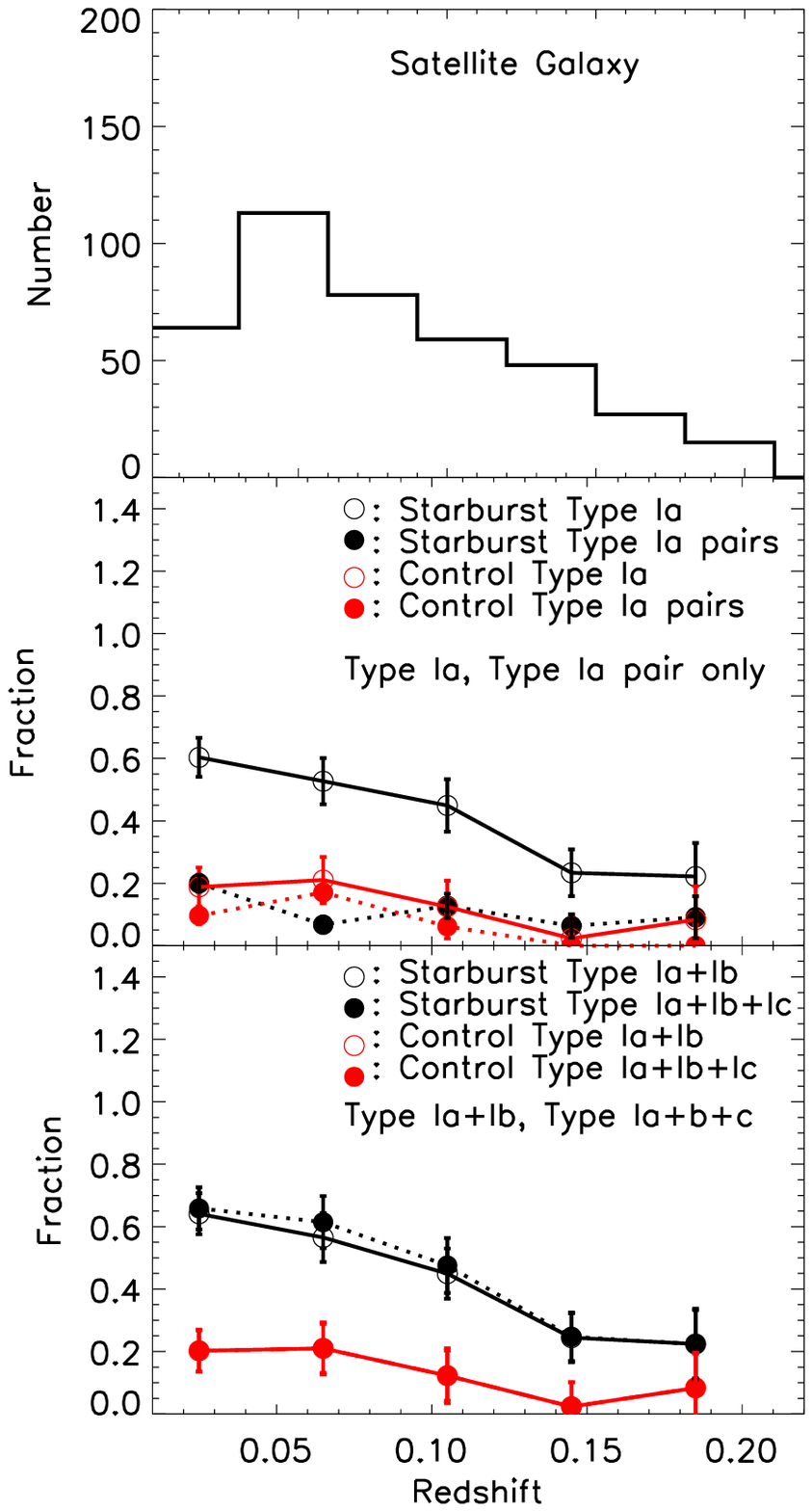}
\includegraphics[width=8cm,height=15cm]{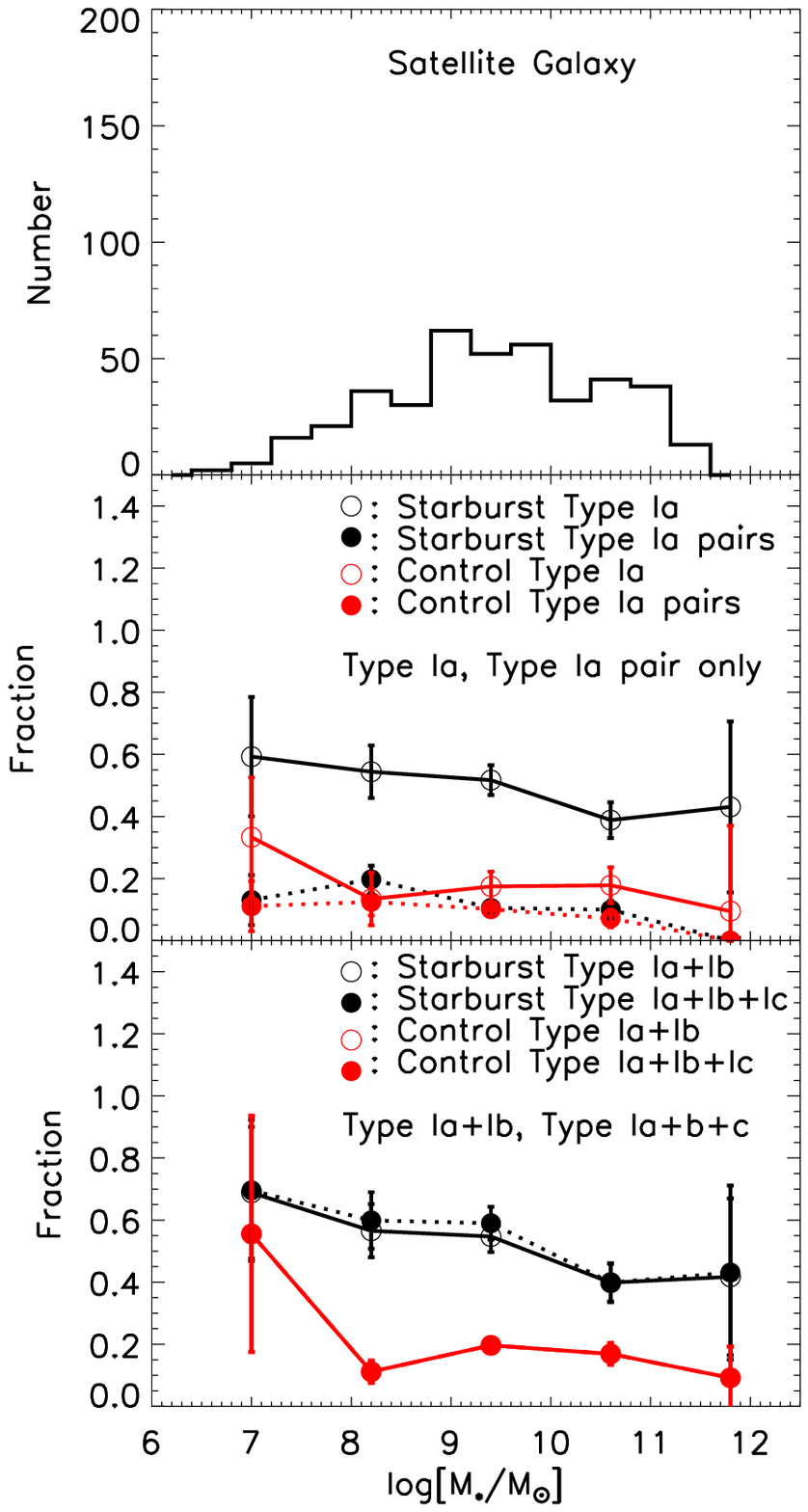}
\caption{Same as Figure \ref{fig:cent} for satellite
  galaxies. }\label{fig:sate}
\end{figure*}

\section{Results}

After classifying all galaxies in our starburst and normal star
forming samples in the previous section, we proceed to study their
relative fractions in different types.

First, we list in Table \ref{tab:frac} the overall fractions of the
different classification groups in both starburst and control
samples. For the starburst galaxy sample, there are no big
differences between centrals and satellites. In total, there are
$\sim$50\% central and 49\% satellite starburst galaxies classified
as Type Ia.  Type Ia+Ib+Ic together reaches $55.3\pm 0.5\%$ for
central galaxies and $54.0\pm 2.4\%$ for satellite galaxies.  Here
the errors are calculated using 200 bootstrap re-samplings.  More
precisely, we have 736 central galaxies classified as Type Ias.
Within these Type Ias, 155 have both spectroscopic companions and
tidal features, 97 only have spectroscopic companions, and the other
484 have only tidal features. For satellites, we have 197 Type Ias,
among which 35 have both close companions and tidal features, 18
have only close companions, and 144 with only tidal features.
Compared to the starburst galaxies, the normal star forming galaxies
show significantly reduced merger signals, i.e. the fraction of Type
Ia is a factor of 2.5 lower, $\sim$19\% in centrals, and 17\% in
satellites. Including the possible merger associated Type Ib and Ic
galaxies, the related fractions increase only by $\sim$2\% and 1\%
for central and satellite galaxies, which is also $\sim$2.5 times
lower.

We next proceed to check the stellar mass ratio of our merger
associated galaxies in Type Ia+Ib+Ic.  As pointed out in Cox et al.
(2008), based on numerical simulations, only major mergers (galaxies
with comparable masses, i.e., with stellar mass ratio within 1:3 and
3:1) can significantly enhance the SFRs of merging galaxies, i.e.,
trigger starburst.  In Type Ias, the major or minor merger fractions
are determined only for galaxies with close companions. For the Type
Ib+Ic, we make use of the photometric redshift together with the
$z$-band apparent magnitude to estimate the stellar mass using the
average mass-to-light ratio provided in Kauffmann et al. (2003). The
resulting values for major merger events are also listed in Table
\ref{tab:frac}. Among our Type Ias with close companions, 63.2\%
(91.1\%) central and 65.5\% (57.9\%) satellite galaxies in our
starburst (control) sample are major merger associated galaxies. If
we assume all the Type Ias without close companions are major merger
associated galaxies, as only a major merger will induce significant
morphology changes (e.g. Mo et al. 2010), the total fractions of the
major mergers are $\sim$45\% and 47\% in starburst central and
satellite galaxy samples. In contrast, these numbers are $\sim$19\%
and 15\% in control samples.  Note however, as the total number of
star forming galaxies is significantly larger than that of the
starburst galaxies, most of the major mergers are indeed associated
with normal star forming galaxies, i.e., a major merger does not
always trigger a starburst. This is consistent with the simulation
results obtained by Di Matteo et al. (2007) that only a small
fraction of mergers trigger starburst and by Cox et al. (2008) that
starburst only lasts a short period of time.

In addition to the major/minor merger separation, we also check what
fractions of the observed mergers are wet-wet or wet-dry mergers.
Here wet galaxies correspond to star forming or blue galaxies
depending on whether or not SFR information is available. In
contrast, dry galaxies correspond to quenched or red galaxies. For
those galaxies without SFR measurements, we separate them in to blue
(wet) and red (dry) using their $g-r$ colors as a function of
luminosity according to the criteria  given in Yang et al. (2008),
\begin{equation}
 ^{0.1} (g-r)=1.022-0.0651x-0.0031x^2\,,
\label{eq:color}
\end{equation}
where $x=\rmag+23.0$ and $^{0.1}{\rm M}_r$ is the absolute magnitude
in the $r$ band K+E corrected to z=0.1.  For Type Ia, the wet-wet
merger fractions are also determined only for galaxies with
spectroscopic close companions. We list in Table \ref{tab:frac} the
related fractions that can be specified into wet-wet mergers. Within
the total Type Ias with close companions, $\sim$95\% in the
starburst sample and 90\% in the control sample are wet-wet mergers.
Such a high fraction is consistent with the "galactic conformity"
found by Weinmann et al. (2006) that satellite galaxies tend to have
the same types (early or late) as their primary galaxy.

Finally, we check if the total fractions have any dependence on the
redshift or the stellar mass of the galaxies in consideration. We
first focus on the central galaxies.  Shown in the upper panels of
Figure  \ref{fig:cent} are the number distribution of galaxies as a
function of redshift (left panel) and stellar mass (right panel).
The distributions of galaxies shown in these two panels reflect the
selection effects of the survey and are referenced for further
fraction measurements.  We divided our starburst galaxy samples into
five redshifts bins and five stellar mass bins.  Shown in the middle
panels of Figure \ref{fig:cent} are results for the Type Ias (open
circles) and Type Ias with close companions only (solid dots).
Compared to the control sample, the starburst sample shows overall
higher amplitudes and slightly decreasing trends both as a function
of redshift and as a function of stellar mass.
The lower fraction of mergers in higher redshift and most massive
starburst galaxies suggests that the starburst in these galaxies are
not necessarily triggered by major mergers, but instead by minor
mergers or accretion of gas from low-mass galaxies. In the lower
panels of Figure  \ref{fig:cent}, we show the results for Type Ia+Ib
and Ia+Ib+Ic. The overall trends are quite similar to those of Type
Ias. Then for the satellite galaxies, similar analysis are carried
out and shown in Figure \ref{fig:sate}. The overall dependences of
satellite galaxies in both our starburst and control samples are
quite consistent with those of central galaxies. The general
consistent behaviors between central and satellite galaxies suggests
that the starbursts are very local events and not much affected by
their large scale environment.

\section{Conclusions}

In this Letter, we report our investigation on the interaction rate
of "starburst" galaxies selected from the SDSS DR7, with respect to
a control sample of normal star forming galaxies. Here the starburst
galaxies are defined with SFRs a factor of five larger than the
median SFR of the star forming galaxies distribution as a function
of stellar mass. We study the fractions of these starburst galaxies
that can be clearly or possibly classified as merger associated. Our
classification is based on visual inspection of all these galaxy
images and pair determination using redshift/distance separation,
separately for central and satellite galaxies. The main conclusions
are as follows.

\begin{enumerate}

\item According to our visual inspection, $\sim$50\% central and 49\%
  satellite starburst galaxies show evident merger features.  In
  contrast, the corresponding number in the control sample of normal
  star forming galaxies falls to $\sim$19\% and 17\%.

\item Taking into account the close companions identified by the
  photometric redshifts, the interaction rates may increase by
  $\sim$5\% in starburst sample and  $\sim$2\% in control sample.

\item More than $\sim$60\% ($\sim$90\%) evident mergers with close
  companions are major (wet-wet) merger associated galaxies. Assuming
  that the starburst galaxies without close companions
  but with disturbed features are all associated with major and
  wet-wet mergers, the total major and wet-wet merger fractions in our
  starburst sample are $\sim$45\% for central galaxies and $\sim$47\%
  for satellite galaxies. In contrast, these numbers are $\sim$19\%
  and $\sim$15\% in control samples.

\item Distinction between central and satellite galaxies do not
  strongly affect the merger rates. This suggests that large scale
  environment has little impact on the starbursts which can thus be
  interpreted as local events.

\end{enumerate}

\acknowledgements

We thank the anonymous referee for helpful comments that greatly
improved this Letter.  We thank Dylan Tweed for useful discussions.
This work is supported by the grants from NSFC (Nos. 10925314,
11128306, 11121062, 11233005) and the Strategic Priority Research
Program "The Emergence of Cosmological Structures" of the Chinese
Academy of Sciences, Grant No. XDB09000000.


\begin{thebibliography}{}

\bibitem[Abazajian et al.(2009)]{2009ApJS..182..543A} Abazajian, K.~N.,
Adelman-McCarthy, J.~K., Ag{\"u}eros, M.~A., et al.\ 2009, \apjs, 182, 543


\bibitem[Arp(1995)]{1995yCat.7074....0A} Arp, H.~C.\ 1995, yCat, 7074, 0


\bibitem[Blanton et al.(2005)]{2005AJ....129.2562B} Blanton, M.~R.,
  Schlegel, D.~J., Strauss, M.~A., et al.\ 2005, \aj, 129, 2562


\bibitem[Bolton et al.(2012)]{2012AJ....144..144B} Bolton, A.~S., Schlegel,
  D.~J., Aubourg, {\'E}., et al.\ 2012, \aj, 144, 144


\bibitem[Bridge et al.(2007)]{2007ApJ...659..931B} Bridge, C.~R.,
  Appleton, P.~N., Conselice, C.~J., et al.\ 2007, \apj, 659, 931


\bibitem[Brinchmann et al.(2004)]{2004MNRAS.351.1151B} Brinchmann, J.,
  Charlot, S., White, S.~D.~M., et al.\ 2004, \mnras, 351, 1151

\bibitem[Bushouse(1987)]{1987ApJ...320...49B} Bushouse, H.~A.\ 1987, \apj,
320, 49


\bibitem[Carliles et al.(2010)]{2010ApJ...712..511C} Carliles, S.,
  Budav{\'a}ri, T., Heinis, S., Priebe, C., \& Szalay, A.~S.\ 2010, \apj, 712,
  511

\bibitem[Chen et al.(2010)]{2010ApJ...712.1385C} Chen, Y., Lowenthal, J.~D.,
  \& Yun, M.~S.\ 2010, \apj, 712, 1385


\bibitem[Cox et al.(2006)]{2006MNRAS.373.1013C} Cox, T.~J., Jonsson, P.,
Primack, J.~R., \& Somerville, R.~S.\ 2006, \mnras, 373, 1013


\bibitem[Cox et al.(2008)]{2008MNRAS.384..386C} Cox, T.~J., Jonsson,
  P., Somerville, R.~S., Primack, J.~R., \& Dekel, A.\ 2008, \mnras,
  384, 386

\bibitem[Csabai et al.(2003)]{2003AJ....125..580C} Csabai, I.,
  Budav{\'a}ri, T., Connolly, A.~J., et al.\ 2003, \aj, 125, 580


\bibitem[Cui et al.(2001)]{2001AJ....122...63C} Cui, J., Xia, X.-Y.,
  Deng, Z.-G., Mao, S., \& Zou, Z.-L.\ 2001, \aj, 122, 63


\bibitem[Di Matteo et al.(2007)]{2007A&A...468...61D} Di Matteo, P.,
  Combes, F., Melchior, A.-L., \& Semelin, B.\ 2007, \aap, 468, 61


\bibitem[Di Matteo et al.(2008)]{2008A&A...492...31D} Di Matteo, P.,
  Bournaud, F., Martig, M., et al.\ 2008, \aap, 492, 31

\bibitem[Elmegreen(2011)]{2011EAS....51...45E} Elmegreen, B.~G.\ 2011, EAS
Publications Series, 51, 45


\bibitem[Kapferer et al.(2005)]{2005A&A...438...87K} Kapferer, W., Knapp, A.,
  Schindler, S., Kimeswenger, S., \& van Kampen, E.\ 2005, \aap, 438, 87

\bibitem[Kartaltepe et al.(2012)]{2012ApJ...757...23K} Kartaltepe, J.~S.,
Dickinson, M., Alexander, D.~M., et al.\ 2012, \apj, 757, 23


\bibitem[Kauffmann et al.(2003)]{2003MNRAS.341...33K} Kauffmann, G., Heckman,
  T.~M., White, S.~D.~M., et al.\ 2003, \mnras, 341, 33

\bibitem[Kennicutt(1998)]{1998ApJ...498..541K} Kennicutt, R.~C., Jr.\ 1998,
\apj, 498, 541

\bibitem[Kewley et al.(2001)]{2001ApJ...556..121K} Kewley, L.~J., Dopita,
  M.~A., Sutherland, R.~S., Heisler, C.~A., \& Trevena, J.\ 2001, \apj, 556,
  121


\bibitem[Larson \& Tinsley(1978)]{1978ApJ...219...46L} Larson, R.~B., \&
  Tinsley, B.~M.\ 1978, \apj, 219, 46

\bibitem[Lee et al.(2007)]{2007ApJ...671L.113L} Lee, J.~C., Kennicutt, R.~C.,
  Funes, S.~J., Jos{\'e} G., Sakai, S., \& Akiyama, S.\ 2007, \apjl, 671, L113


\bibitem[Lee et al.(2009)]{2009ApJ...692.1305L} Lee, J.~C., Kennicutt, R.~C.,
  Jr., Funes, S.~J.~J.~G., Sakai, S., \& Akiyama, S.\ 2009, \apj, 692, 1305


\bibitem[Lupton et al.(2002)]{2002SPIE.4836..350L} Lupton, R.~H.,
  Ivezic, Z., Gunn, J.~E., et al.\ 2002, \procspie, 4836, 350

\bibitem[Lupton et al.(2004)]{2004PASP..116..133L} Lupton, R., Blanton,
M.~R., Fekete, G., et al.\ 2004, \pasp, 116, 133


\bibitem[Melnick \& Mirabel(1990)]{1990A&A...231L..19M} Melnick, J.,
  \& Mirabel, I.~F.\ 1990, \aap, 231, L19

\bibitem[Mesa et al.(2013)]{2013arXiv1312.0560M} Mesa, V., Duplancic, F.,
Alonso, S., Coldwell, G., \& Lambas, D.~G.\ 2014, MNRAS, 438, 1784


\bibitem[Meza et al.(2003)]{2003ApJ...590..619M} Meza, A., Navarro, J.~F.,
Steinmetz, M., \& Eke, V.~R.\ 2003, \apj, 590, 619

\bibitem[Mihos \& Hernquist(1994)]{1994ApJ...425L..13M} Mihos, J.~C., \&
  Hernquist, L.\ 1994, \apjl, 425, L13

\bibitem[Mo et al.(2010)]{2010gfe..book.....M} Mo, H., van den Bosch, F.~C.,
  \& White, S.\ 2010, Galaxy Formation and Evolution.~Cambridge University
  Press, 2010.~ISBN: 9780521857932


\bibitem[Murphy et al.(1996)]{1996AJ....111.1025M} Murphy, T.~W., Jr.,
  Armus, L., Matthews, K., et al.\ 1996, \aj, 111, 1025


\bibitem[Patton et al.(2000)]{2000ApJ...536..153P} Patton, D.~R.,
  Carlberg, R.~G., Marzke, R.~O., et al.\ 2000, \apj, 536, 153

\bibitem[Sanders et al.(1988)]{1988ApJ...325...74S} Sanders, D.~B.,
  Soifer, B.~T., Elias, J.~H., et al.\ 1988, \apj, 325, 74

\bibitem[Rodighiero et al.(2011)]{2011ApJ...739L..40R} Rodighiero, G.,
Daddi, E., Baronchelli, I., et al.\ 2011, \apjl, 739, L40


\bibitem[Salim et al.(2007)]{2007ApJS..173..267S} Salim, S., Rich,
  R.~M., Charlot, S., et al.\ 2007, \apjs, 173, 267

\bibitem[Springel(2000)]{2000MNRAS.312..859S} Springel, V.\ 2000,
  \mnras, 312, 859

\bibitem[Telesco et al.(1988)]{1988ApJ...329..174T} Telesco, C.~M.,
Wolstencroft, R.~D., \& Done, C.\ 1988, \apj, 329, 174


\bibitem[Tissera et al.(2002)]{2002MNRAS.333..327T} Tissera, P.~B.,
Dom{\'{\i}}nguez-Tenreiro, R., Scannapieco, C.,
\& S{\'a}iz, A.\ 2002, \mnras, 333, 327

\bibitem[van Dokkum(2005)]{2005AJ....130.2647V} van Dokkum, P.~G.\ 2005,
\aj, 130, 2647

\bibitem[Weinmann et al.(2006)]{2006MNRAS.366....2W} Weinmann, S.~M.,
  van den Bosch, F.~C., Yang, X., \& Mo, H.~J.\ 2006, \mnras, 366, 2




\bibitem[Yang et al.(2007)]{2007ApJ...671..153Y} Yang, X., Mo, H.~J., van
den Bosch, F.~C., et al.\ 2007, \apj, 671, 153

\bibitem[Yang et al.(2008)]{2008ApJ...676..248Y} Yang, X., Mo, H.~J.,
  \& van den Bosch, F.~C.\ 2008, \apj, 676, 248


\bibitem[Yang et al.(2013)]{2013ApJ...770..115Y} Yang, X., Mo, H.~J., van den
  Bosch, F.~C., et al.\ 2013, \apj, 770, 115

\bibitem[York et al.(2000)]{2000AJ....120.1579Y} York, D.~G., Adelman, J.,
Anderson, J.~E., Jr., et al.\ 2000, \aj, 120, 1579

\end{thebibliography}
\end{document}